\documentclass[%
 reprint,
superscriptaddress,
 amsmath,amssymb,
 aps,
pra,
]{revtex4-1}

\usepackage{amsthm}
\usepackage{graphicx}
\usepackage{dcolumn}
\usepackage{bm}
\usepackage{multirow} 
\usepackage{enumerate}
\usepackage{algorithm}
\usepackage{algorithmic}
\usepackage{tabularx}

\DeclareMathOperator{\supp}{supp}
\DeclareMathOperator{\conf}{conf}

\theoremstyle{remark}

\begin{document}

\makeatletter
\newcommand{\rmnum}[1]{\romannumeral #1}
\newcommand{\Rmnum}[1]{\expandafter\@slowromancap\romannumeral #1@}
\makeatother

\preprint{APS/123-QED}

\title{Quantum algorithm for association rules mining}

\author{Chao-Hua Yu}
\affiliation{State Key Laboratory of Networking and Switching Technology, Beijing University of Posts and Telecommunications, Beijing, 100876, China}
\affiliation{State Key Laboratory of Cryptology, P.O. Box 5159, Beijing, 100878, China}
\author{Fei Gao}
\email{gaof@bupt.edu.cn}
\author{Qing-Le Wang}
\affiliation{State Key Laboratory of Networking and Switching Technology, Beijing University of Posts and Telecommunications, Beijing, 100876, China}
\author{Qiao-Yan Wen}
\affiliation{State Key Laboratory of Networking and Switching Technology, Beijing University of Posts and Telecommunications, Beijing, 100876, China}

\date{\today}

\begin{abstract}
Association rules mining (ARM) is one of the most important problems in knowledge discovery and data mining. Given a transaction database that has a large number of transactions and items, the task of ARM is to acquire consumption habits of customers by discovering the relationships between itemsets (sets of items). In this paper, we address ARM in the quantum settings and propose a quantum algorithm for the key part of ARM, finding out frequent itemsets from the candidate itemsets and acquiring their supports. Specifically, for the case in which there are $M_f^{(k)}$ frequent $k$-itemsets in the $M_c^{(k)}$ candidate $k$-itemsets ($M_f^{(k)} \leq M_c^{(k)}$), our algorithm can efficiently mine these frequent $k$-itemsets and estimate their supports by using parallel amplitude estimation and amplitude amplification with complexity $\mathcal{O}(\frac{k\sqrt{M_c^{(k)}M_f^{(k)}}}{\epsilon})$, where $\epsilon$ is the error for estimating the supports. Compared with the classical counterpart, classical sampling-based algorithm, whose complexity is $\mathcal{O}(\frac{kM_c^{(k)}}{\epsilon^2})$, our quantum algorithm quadratically improves the dependence on both $\epsilon$ and $M_c^{(k)}$ in the best case when $M_f^{(k)}\ll M_c^{(k)}$ and on $\epsilon$ alone in the worst case when $M_f^{(k)}\approx M_c^{(k)}$.


\begin{description}
\item[PACS numbers]
03.67.Dd, 03.67.Hk
\end{description}
\end{abstract}

\pacs{Valid PACS appear here}
\maketitle


\section{Introduction}

Quantum computing provides a paradigm that makes use of quantum mechanical principles, such as superposition and entanglement, to perform computing tasks in quantum systems (quantum computers) \cite{QCQI}. Just as classical algorithms run in the classical computers, a quantum algorithm is a step-by-step procedure run in the quantum computers for solving a certain problem, which, more interestingly, is expected to outperform the classical algorithms for the same problem. As of now, various quantum algorithms have been put forward to solve a number of problems faster than their classical counterparts \cite{QA}, and mainly fall into one of three classes \cite{PWShor}. The first class is based on the quantum Fourier transformation \cite{QCQI}, the most famous representative being Shor's algorithm \cite{Shor} for large number factoring and discrete logarithm, which offers exponential speedup over the classical algorithms. The second class is represented by the Grover's quantum search \cite{Grover} and its generalized version, i.e., amplitude amplification \cite{AA}, both of which achieve quadratic speedup over the classical search algorithm. The third class contains the algorithms for quantum simulation \cite{QS}, the original idea of which is suggested by Feynman \cite{F} to speed up the simulation of quantum systems using quantum computers. In the past decade, quantum simulation has made great progress in efficient sparse Hamiltonian simulation \cite{BCK}.

However, it is a pity that no more fundamental quantum algorithms except the above three classes of quantum algorithms have ever been found. In addition to seeking new algorithms, another important direction for quantum computing seems to apply known quantum algorithms to new problem areas, such as machine learning \cite{QML}.

In the last one and a half decade, quantum machine learning \cite{QML} has become a booming research field and lots of quantum algorithms related to various machine learning problems have been proposed, such as quantum algorithm for solving linear equations \cite{HHL}, quantum linear regression \cite{WBL,SSP}, quantum principal component analysis \cite{QPCA}, quantum supervised learning (data classification) \cite{ARRZ,LMR,RML,CD}, quantum unsupervised learning (data clustering analysis) \cite{ABG,RML}, quantum search engine ranking \cite{GZL,PM,SDGZ,PMCM}, quantum neural network \cite{QNN}, and so on. More excitingly, these algorithms in some degree are shown faster than their classical counterparts. For example, under the condition that the quantum data as inputs are provided, the quantum support vector machine for big data classification exhibits exponential speedup over the classical support vector machine \cite{RML}. Furthermore, since machine learning is a crucial tool for data mining which is a computational process of extract valuable information from a large data set, one of the most important applications of quantum machine learning is to efficiently implement data mining tasks in quantum computers \cite{QDM}.

In this paper, we address association rules mining (ARM)\cite{DM}, one of the most important problems in big data mining, in the quantum settings.
Given a transaction database consisting of a large number of transactions and items, the task of ARM is to discover the association rules connecting two itemsets (an itemset is a set of items) $A$ and $B$ in the conditional implication form $A \Rightarrow B$, which implies that a customer who buys the items in $A$ also tends to buy the items in $B$. The core of ARM is to mine the itemsets that frequently occur in the transactions, which entails finding out the itemsets whose supports (occurrence frequency) are not less than a pre-specified threshold, i.e., frequent itemsets, from a number of candidate itemsets \cite{DM,Apriori}. Herein we provide an efficient quantum algorithm for ARM based on the oracle accessing the database. In particular, for mining frequent $k$-itemsets (a $k$-itemset is a set of $k$ items) from candidate $k$-itemsets, we first perform parallel amplitude estimation to estimate the supports of all the candidate $k$-itemsets. In other words, a quantum superposition state with each superposed term encoding the support of a candidate $k$-itemset is generated. After that, by employing the amplitude amplification, we then search in the state for the candidate $k$-itemsets with supports not less than the thereshold. We analyze the query complexity of our algorithm and it is shown that, compared with the classical counterpart, sampling-based algorithm \cite{MTV}, our algorithm improves the complexity at least in the dependence on the error of estimating the supports in amplitude estimation while keeping the other parameters invariant.

The rest of this paper is organized as follows. In Sec. \ref{sec:2}, we review ARM in terms of its basic concepts, notations and classical algorithmic procedures. Sec. \ref{sec:3} presents the details of our quantum algorithm and gives complexity analysis on this algorithm.  Discussions and conclusions are given in the last section.

\section{Review of ARM}
\label{sec:2}
In this section, we briefly review some basic concepts and notations of ARM and classical algorithmic procedures implementing ARM. More details can be seen in the reference \cite{DM}.

\begin{figure}[b]
\includegraphics[scale=0.5]{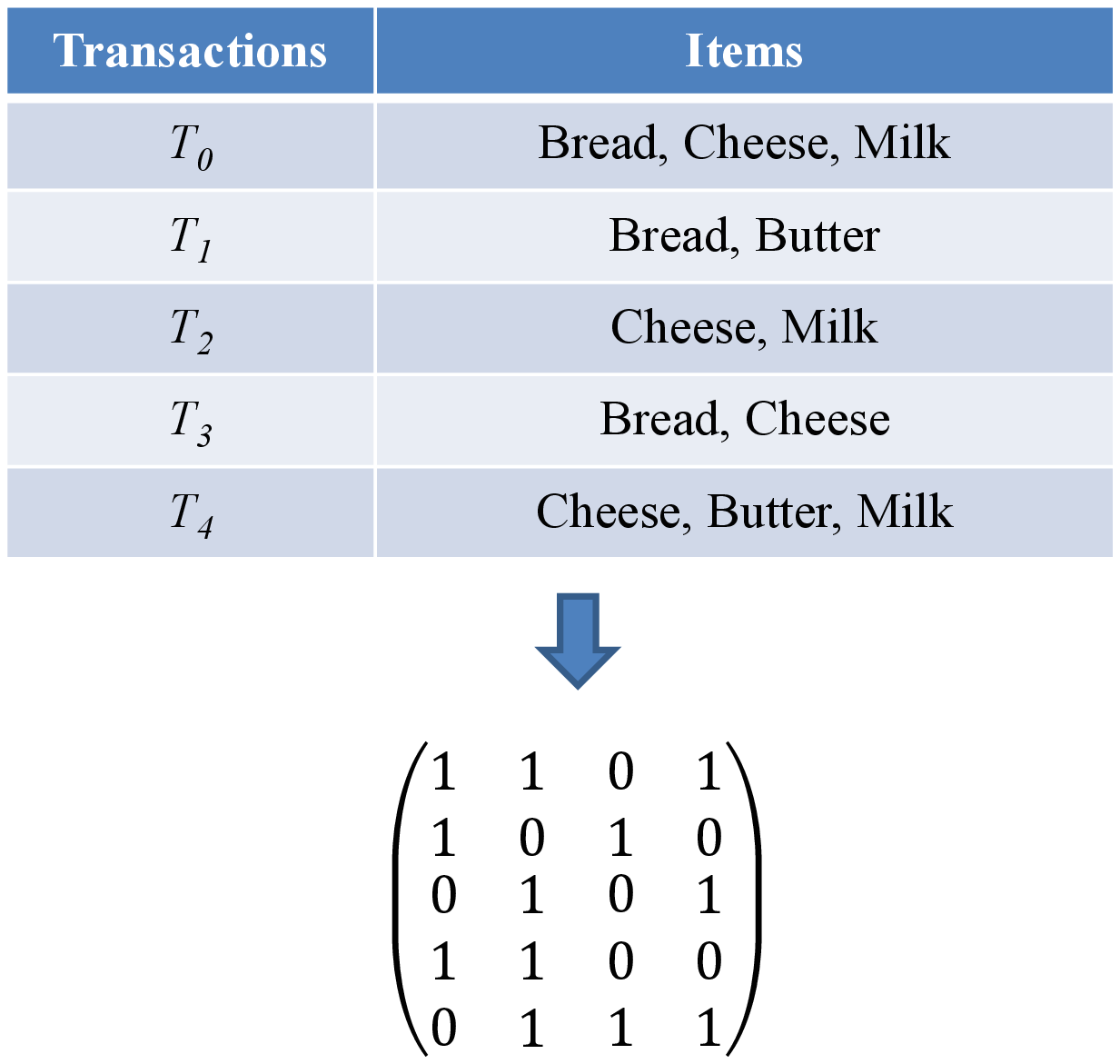}
\caption{\label{fig1} An example of transaction database that contains five transactions $\mathcal{T}=\{T_0,T_1,T_2,T_3,T_4\}$ with each one being a subset of the set of four items $\mathcal{I}=\{I_0= Bread,I_1=Cheese,I_2=Butter,I_3=Milk\}$ and its binary matrix representation.}
\end{figure}

A transaction database, the objective ARM deals with, that contains $N$ transactions can be denoted by the set $\mathcal{T}=\{T_0,T_1,\cdots,T_{N-1} \}$ and each transaction is a subset of the set of $M$ items $\mathcal{I}=\{I_0,I_1,\cdots,I_{M-1} \}$, i.e., $T_i \subseteq \mathcal{I}$. It can also be represented by a $N \times M$ binary matrix, denoted by $D$, in which the element $D_{ij}=1 (0)$ means that the item $I_j$ is (not) contained in the transaction $T_i$. To illustrate it, a simple example is given in the Fig. \ref{fig1}.

A set of items is called an \emph{itemset}. The \emph{support} of an itemset $X$ is defined as the proportion of transactions in $T$ that contain all the items in $X$, i.e., $\supp(X)=\frac{|\{T_i| X \subseteq T_i\}|}{N}$. An association rule is of the implication form $A \Rightarrow B$, where $A$ and $B$ are two disjoint itemsets.
Its support is defined as $\supp(A \Rightarrow B)=\supp(A\cup B)$ and its confidence is defined as $\conf(A \Rightarrow B)=\frac{\supp(A\cup B)}{\supp(A)}$. A rule is called frequent (confident) if its support (confidence) is not less than a prespecified threshold $min\_supp$ ($min\_conf$). The task of ARM is to find out the rules $A \Rightarrow B$ that are both frequent and confident. Implementing this task is consist of two phases \cite{DM}:\\
\indent (1) find out all the frequent itemsets $X$, defined as $\supp(X) \geq min\_supp$;\\
\indent (2) find out all the confident rules $A \Rightarrow B$ such that $A\cup B=X$.

\begin{figure}[b]
\includegraphics[scale=0.4]{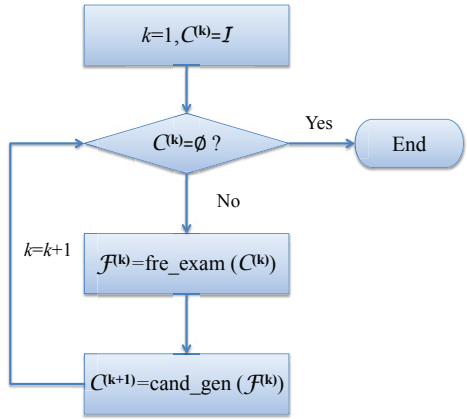}
\caption{\label{fig2} The whole process of Apriori algorithm.}
\end{figure}

Since the second phase is much less costly than the first \cite{DM}, the core work of ARM lies in the first phase. Therefore, the task of mining association rules can be reduced to that of mining frequent itemsets.
In classical regime, there are various algorithms \cite{DM} for mining frequent itemsets, the most famous one being the \emph{Apriori} algorithm \cite{DM,Apriori}. Based on the important \emph{Apriori property} stating that all nonempty subset of a frequent itemset must also be frequent, Apriori algorithm employs an iterative approach known as a level-wise search to discover all the frequent itemsets, the whole process of which is depicted in the Fig. \ref{fig2}. In the $k$th iteration of the algorithm, two procedures are executed: \\
\indent (P1) Given the set of \emph{candidate} $k$-itemsets $\mathcal{C}^{(k)}$ which is just $\mathcal{I}$ when $k=1$, the supports of all the elements in $\mathcal{C}^{(k)}$ are examined by passing every transaction of database and the frequent elements are picked out to form the set of all frequent $k$-itemsets $\mathcal{F}^{(k)}$. This procedure can be seen as performing a function $\textbf{fre\_exam}$ that finds out frequent itemsets from candidate itemsets, namely $\mathcal{F}^{(k)}=\textbf{fre\_exam}(\mathcal{C}^{(k)})$. \\
\indent (P2) Generate the set of candidate $(k+1)$-itemsets $\mathcal{C}^{(k+1)}$ from $\mathcal{F}^{(k)}$. This procedure is generally consist of two steps, join step and prune step \cite{DM}, and can also be seen as performing a function $\textbf{cand\_gen}$, namely $\mathcal{C}^{(k+1)}=\textbf{cand\_gen}(\mathcal{F}^{(k)})$.

In practice, in each iteration (P1) dominates the time complexity of the whole process \cite{MTV}. Therefore, how to efficiently execute (P1) of each iteration, namely finding out frequent itemsets from candidate ones, is of great importance. In the following section, we provide a quantum algorithm
to implement (P1) for each iteration that can significantly reduce the time complexity in contrast to the classical algorithms.

\section{Quantum Algorithm for ARM}
\label{sec:3}
In this section, we are to design a quantum algorithm based on the basic oracle $O$ that can access the elements of the database binary matrix $D$. In the first place, we show how to use the basic oracle to construct oracles $O^{(k)}$ that can identify whether a transaction contain an $k$-itemset. Then we in detail present our algorithm that takes $O^{(k)}$ as the elementary subroutine. Finally, we analyze the query complexity of our algorithm and compare it with that of classical algorithms.

\subsection{Constructing the oracles $O^{(k)}$ by using the basic oracle $O$}
\begin{figure}[b]
\includegraphics[scale=0.5]{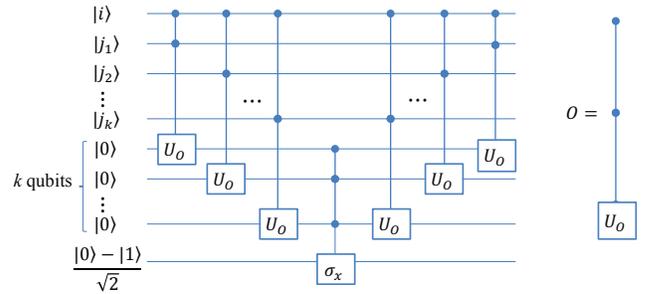}
\caption{\label{fig3} The left part is the quantum circuit for constructing the oracle $O^{(k)}$ by using the basic oracle $O$ and the generalized CNOT operation $\bigwedge_k(\sigma_x)$, where the circuit representation of $O$ is given in the right part.}
\end{figure}

In our algorithm, the basic oracle $O$ is precisely an unitary operation acting on the computational basis,
\begin{eqnarray}
\label{Eq:1}
O|i\rangle|j\rangle|a\rangle = |i\rangle|j\rangle|a \oplus D_{ij}\rangle,
\end{eqnarray}
where $i$ ranges in $\mathbb{Z}_N$ and $j$ ranges in $\mathbb{Z}_M$. Just as the standard Grover's algorithm \cite{QCQI}, by taking $|a\rangle=\frac{|0\rangle-|1\rangle}{\sqrt{2}}$, this oracle can be employed to construct a new oracle $O^{(1)}$ acting as
\begin{eqnarray}
\label{Eq:2}
O^{(1)}|i\rangle|j\rangle = (-1)^{D_{ij}}|i\rangle|j\rangle,
\end{eqnarray}
which flip the phase of the state $|i\rangle|j\rangle$ when the transaction $T_i$ contains the item $I_j$ (i.e., $D_{ij}=1$). Furthermore, $O$ can be applied as the primitive to construct more complex oracles $O^{(k)}$ that can identify whether a transaction contains an $k$-itemset  $X=\{I_{j_l}|l=1,2,\cdots,k\}$ acting as
\begin{eqnarray}
\label{Eq:3}
O^{(k)}|i\rangle|j_1\rangle|j_2\rangle\cdots|j_k\rangle = (-1)^{\tau(i,X)}|i\rangle|j_1\rangle|j_2\rangle\cdots|j_k\rangle,
\end{eqnarray}
where $\tau(i,X)=\prod_{l=1}^{k}D_{ij_l}$ is a boolean value which identifies whether the transaction $T_i$ contains $X$, or equivalently, whether $X \subseteq T_i$.
That is, if $X \subseteq T_i$ (i.e., $\tau(i,X)=1$), the phase of the state $|i\rangle|j_1\rangle|j_2\rangle\cdots|j_k\rangle$ would be flipped; otherwise, the phase would not be affected. Construction of $O^{(k)}$ requires the basic oracle $O$ and also the generalized CNOT operation $\bigwedge_k(\sigma_x)$ ($\sigma_x$ is the pauli matrix \cite{QCQI}) which use $\Theta(k)$ \cite{Complexity} basic one-qubit and two-qubit gates to carry out the map \cite{Gate}
\begin{eqnarray}
\label{Eq:4}
|x_1\rangle|x_2\rangle\cdots|x_k\rangle|y\rangle \mapsto |x_1\rangle|x_2\rangle\cdots|x_k\rangle|y\oplus \prod_{i=1}^k x_i\rangle.
\end{eqnarray}
The detailed process of the construction can be illustrated by the quantum circuit shown in Fig. \ref{fig3} which is in fact consist of the following steps: \\
\indent (1) prepare four registers in the state $|i\rangle (|j_1\rangle|j_2\rangle\cdots|j_k\rangle) ( \overbrace{|0\rangle|0\rangle\cdots|0\rangle}^k ) \frac{|0\rangle-|1\rangle}{\sqrt{2}}$;\\
\indent (2) perform the operation $O_k O_{k-1}\cdots O_1$ on the state and then obtain the state $|i\rangle |j_1\rangle|j_2\rangle\cdots|j_k\rangle |D_{ij_1}\rangle|D_{ij_2}\rangle\cdots|D_{ij_k}\rangle\frac{|0\rangle-|1\rangle}{\sqrt{2}}$, where $O_l$ is the operation of performing the oracle $O$ on $|i\rangle$, $|j_l\rangle$ and the $l$th $|0\rangle$;\\
\indent (3) apply the operation $\bigwedge_k(\sigma_x)$ to the last $k+1$ qubits and we have the state $(-1)^{\prod_{l=1}^k D_{ij_1}}|i\rangle |j_1\rangle|j_2\rangle\cdots|j_k\rangle |D_{ij_1}\rangle|D_{ij_2}\rangle\cdots|D_{ij_k}\rangle\frac{|0\rangle-|1\rangle}{\sqrt{2}}$;
\indent (4) reverse the step (2), discard the last $k+1$ qubits and then the oracle $O^{(k)}$ is implemented as Eq. (\ref{Eq:3}).

From the above process, it is easy to see construction of $O^{(k)}$ requires $2k$ basic oracles $O$ and $\Theta(k)$ basic one-qubit or two-qubit gates.

\subsection{Algorithm}
Now we use the oracle $O^{(k)}$ to design our ARM algorithm to mine $\mathcal{F}^{(k)}$ from $\mathcal{C}^{(k)}$. Schematically, our algorithm will firstly estimate the supports of all the candidate $k$-itemsets (elements) in $\mathcal{C}^{(k)}$  in parallel by using amplitude estimation, and then search for candidate $k$-itemsets with supports not less than $min\_supp$ by employing amplitude amplification to obtain the the set of frequent $k$-itemsets $\mathcal{F}^{(k)}$. Here we suppose $\mathcal{C}^{(k)}$ has $M_c^{(k)}$ elements $\mathcal{C}^{(k)}=\{C_j^{(k)}|j=1,2,\cdots,M_c^{(k)}\}$ where $C_j^{(k)}=\{I_{c_{jl}^{(k)}}|l=1,2,\cdots,k,c_{jl}^{(k)} \in \mathbb{Z}_M\}$, $\mathcal{F}^{(k)}$ has $M_f^{(k)}$ elements and $\mathcal{F}^{(k)} \subseteq \mathcal{C}^{(k)}$.
Mining frequent $k$-itemsets from $\mathcal{C}^{(k)}$ in the first place entails acquiring the supports of all the candidate $k$-itemsets $C_j^{(k)}$ in $\mathcal{C}^{(k)}$. Here we denote the support of $C_j^{(k)}$ by $s_j^{(k)}$. For a particular candidate $k$-itemset $C_j^{(k)}$, a direct method for estimating its support $s_j^{(k)}$ in a quantum computer would be the use of amplitude estimation \cite{AA}.

Now we give a brief description of how quantum amplitude estimation works for estimating $s_j^{(k)}$. To achieve this task, a related oracle denoted by $O_j^{(k)}$ that should act as
\begin{eqnarray}
\label{Eq:5}
O_j^{(k)}|i\rangle=(-1)^{\tau(i,C_j^{(k)})}|i\rangle
\end{eqnarray}
is required and its corresponding Grover operator is
\begin{eqnarray}
\label{Eq:6}
G_j^{(k)}=(2|\mathcal{X}_N\rangle\langle\mathcal{X}_N|-\mathbb{I}_N)O_j^{(k)},
\end{eqnarray}
where $|\mathcal{X}_N\rangle := \frac{\sum_{i=0}^{N-1}|i\rangle}{\sqrt{N}}$ and $\mathbb{I}_N$ is the identity matrix with dimension $N$. $G_j^{(k)}$ has two eigenvalues $\lambda_{\pm}=e^{\pm 2\iota \theta_j^{(k)}}$ ($\iota=\sqrt{-1}$ denotes the principal square root of $-1$) and corresponding eigenvectors $|\phi_{j\pm}^{(k)}\rangle$. As a matter of fact,
\begin{eqnarray}
\label{Eq:7}
s_j^{(k)}=\sin^2(\theta_j^{(k)}).
\end{eqnarray}
If we initialize two registers in the state $(\frac{\sum_{t=0}^{T-1}|t\rangle}{\sqrt{T}})|\mathcal{X}_N\rangle$
and take $G_j^{(k)}$ as Grover operator to perform amplitude (phase) estimation \cite{AA} on the state, we will finally attain the state
\begin{eqnarray}
\label{Eq:8}
|\Phi_j^{(k)}\rangle&&=\frac{e^{\iota\theta_j^{(k)}}}{\sqrt{2}}|\mathcal{E}_T(\frac{\theta_j^{(k)}}{\pi})\rangle|\phi_{j+}^{(k)}\rangle \nonumber\\
&&-\frac{e^{-\iota\theta_j^{(k)}}}{\sqrt{2}}|\mathcal{E}_T(1-\frac{\theta_j^{(k)}}{\pi})\rangle|\phi_{j-}^{(k)}\rangle
\end{eqnarray}
where the global phase is ignored and $|\mathcal{E}_T(\omega)\rangle=|T\omega\rangle$ when $T\omega$ is an integer and otherwise
\begin{eqnarray}
\label{Eq:9}
|\mathcal{E}_T(\omega)\rangle = \sum_{y=0}^{T-1} \frac{e^{2\pi \iota (T\omega-y)}-1}
{T(
e^{\frac{2\pi \iota (T\omega-y)}{T}}-1
)}|y\rangle.
\end{eqnarray}
Then measuring $|\Phi_j^{(k)}\rangle$ in the computational basis in the first register will with a high probability output some $\tilde{y}_j$ or $T-\tilde{y}_j$ such that $\sin^2(\frac{\pi\tilde{y}_j}{T})=\sin^2(\frac{\pi(T-\tilde{y}_j)}{T})\sim s_j^{(k)}$; thus $\sin^2(\frac{\pi\tilde{y}_j}{T})$ or $\sin^2(\frac{\pi(T-\tilde{y}_j)}{T})$ can be taken as the estimate for $s_j^{(k)}$.

Surprisingly, when confining to $C_j^{(k)}$ and letting
\begin{eqnarray}
\label{Eq:10}
|C_j^{(k)}\rangle :=\otimes_{l=1}^{k}|c_{jl}^{(k)}\rangle,
\end{eqnarray}
$O^{(k)}$ has the same function as $O_j^{(k)}$
(shown in Eq. (\ref{Eq:5})) according to Eq. (\ref{Eq:3}),
\begin{eqnarray}
\label{Eq:11}
O^{(k)}|i\rangle|C_j^{(k)}\rangle &=& (-1)^{\tau(i,C_j^{(k)})}|i\rangle|C_j^{(k)}\rangle \nonumber\\
&=& (O_j^{(k)}|i\rangle)|C_j^{(k)}\rangle.
\end{eqnarray}
Therefore, based on $O^{(k)}$, we have an Grover-like operator
\begin{eqnarray}
\label{Eq:12}
G^{(k)}=\big((2|\mathcal{X}_N\rangle\langle\mathcal{X}_N|-\mathbb{I}_N)\otimes \mathbb{I}_{M^k} \big)O^{(k)}
\end{eqnarray}
in contrast with $G_j^{(k)}$ (Eq. (\ref{Eq:6})), where $\mathbb{I}_{M^k}$ is due to that the dimension of $|C_j^{(k)}\rangle$ is $M^k$. Then, from the Eqs. (\ref{Eq:5}), (\ref{Eq:6}), (\ref{Eq:11}) and (\ref{Eq:12}),  it is easy to derive that for any integer $y>0$ we have
\begin{eqnarray}
\label{Eq:13}
\big(G^{(k)}\big)^y \big( |\mathcal{X}_N\rangle|C_j^{(k)}\rangle \big)=\bigg( \big(G_j^{(k)}\big)^y|\mathcal{X}_N\rangle \bigg)|C_j^{(k)}\rangle.
\end{eqnarray}
Consequently, if we take $G^{(k)}$ to perform amplitude estimation on the three-register state $(\frac{\sum_{t=0}^{T-1}|t\rangle}{\sqrt{T}})|\mathcal{X}_N\rangle|C_j^{(k)}\rangle$ instead of $(\frac{\sum_{t=0}^{T-1}|t\rangle}{\sqrt{T}})|\mathcal{X}_N\rangle$, we can finally get the state $|\Phi_j^{(k)}\rangle|C_j^{(k)}\rangle$ in contrast with $|\Phi_j^{(k)}\rangle$ (Eq. (\ref{Eq:8})). Furthermore, if we take the superposition state $\frac{\sum_{j=1}^{M_c^{(k)}}|C_j^{(k)}\rangle}{\sqrt{M_c^{(k)}}}$ instead of $|C_j^{(k)}\rangle$ as input, we will finally obtain the state
\begin{eqnarray}
\label{Eq:14}
|\Phi^{(k)}\rangle=\frac{\sum_{j=1}^{M_c^{(k)}}|\Phi_j^{(k)}\rangle|C_j^{(k)}\rangle}{\sqrt{M_c^{(k)}}}.
\end{eqnarray}
because of the linearity of of unitary operator. So the estimates of all the supports $s_j^{(k)}$ are stored in the first register in parallel. We call the process in which the "big" Grover-like operator $G^{(k)}$ is taken to perform amplitude estimation \emph{parallel amplitude estimation}.

After parallel amplitude estimation, we perform amplitude amplification on the first register of $|\Phi^{(k)}\rangle$ to search for the terms $y$ such that $\sin^2(\frac{\pi y}{T})\geq min\_supp$ or $\sin^2(\frac{\pi(T-y)}{T})\geq min\_supp$, so that we obtain a superposition state encoding the frequent $k$-itemsets in the third register and their supports in the first register.

The overall process of our quantum algorithm is summarized by the following five steps.\\
\textbf{Algorithm}: $\mathcal{F}^{(k)}$=\textbf{QARM}($\mathcal{C}^{(k)}$,$G^{(k)}$,$k$,$T$) \\
1. Prepare three registers in the state
\begin{eqnarray}
\label{Eq:15}
|\Psi_1\rangle=(\frac{\sum_{t=0}^{T-1}|t\rangle}{\sqrt{T}})|\mathcal{X}_N\rangle (\frac{\sum_{j=1}^{M_c^{(k)}}|C_j^{(k)}\rangle}{\sqrt{M_c^{(k)}}}).
\end{eqnarray}
2. Perform the unitary operation $\sum_{y=0}^{T-1}|y\rangle\langle y|\otimes (G^{(k)})^{y}$ on $|\Psi_1\rangle$ and the result state is
\begin{eqnarray}
\label{Eq:16}
|\Psi_2\rangle=\bigg(\sum_{y=0}^{T-1}|y\rangle\langle y|\otimes (G^{(k)})^{y} \bigg) |\Psi_1\rangle.
\end{eqnarray}
3. Perform the inverse Fourier transformation $F_T^{\dag}$ on the first register of $|\Psi_2\rangle$ and obtain
\begin{eqnarray}
\label{Eq:17}
|\Psi_3\rangle=(F_T^{\dag}\otimes \mathbb{I}_{N} \otimes \mathbb{I}_{M^k})|\Psi_1\rangle=|\Phi^{(k)}\rangle,
\end{eqnarray}
where $F_T$ is defined by $F_T|i\rangle=\sum_{j=0}^{T-1}\frac{e^{\frac{2\pi \iota ij}{T}}|j\rangle}{\sqrt{T}}$.\\
4. Search in the first register of $|\Psi_3\rangle$ for the terms $y$ satisfying $\sin^2(\frac{\pi y}{T})\geq min\_supp$ or $\sin^2(\frac{\pi(T-y)}{T})\geq min\_supp$ by using amplitude amplification and then obtain the state
\begin{eqnarray}
\label{Eq:18}
|\Psi_4\rangle \sim \frac{\sum_{j=1,\supp(C_j^{(k)})\geq min\_supp}^{M_c^{(k)}}|\Phi_j^{(k)}\rangle|C_j^{(k)}\rangle}{\sqrt{M_f^{(k)}}}.
\end{eqnarray}
The state contains three registers holding the estimates of the supports of frequent $k$-itemsets, the eigenstates of the Grover operators $G_j^{(k)}$, and the frequent $k$-itemsets, from left to right, respectively. $|\Phi_j^{(k)}\rangle$ is seen in Eq. (\ref{Eq:8}).\\
5. Measure the first and third register for $\mathcal{O}(M_f^{(k)})$ times to reveal all the $M_f^{(k)}$ frequent $k$-itemsets (i.e., $\mathcal{F}^{(k)}$) and their supports.

\begin{figure}[b]
\includegraphics[scale=0.45]{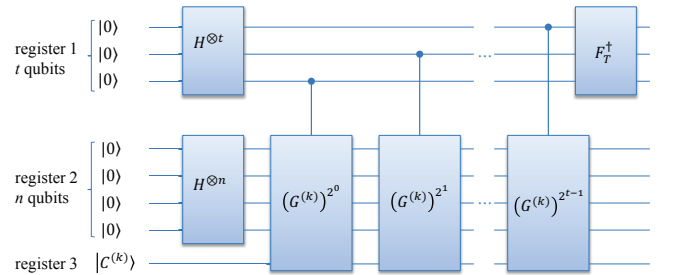}
\caption{\label{fig4} Quantum circuit of the first three steps of our algorithm when $T=2^t$ and $N=2^n$. Here $|C^{(k)}\rangle:=\frac{\sum_{j=1}^{M_c^{(k)}}|C_j^{(k)}\rangle}{\sqrt{M_c^{(k)}}}$.}
\end{figure}

The first three steps contribute to the key part of our algorithm, parallel amplitude estimation, and the circuit for the case that $T$ and $N$ are powers of 2, $T=2^t$ and $N=2^n$, is shown in the Fig. \ref{fig4}.

It should be stressed that, when $k>1$, it is more advisable to replace the superposition state
\begin{eqnarray}
\label{Eq:19}
|C^{(k)}\rangle:=\frac{\sum_{j=1}^{M_c^{(k)}}|C_j^{(k)}\rangle}{\sqrt{M_c^{(k)}}}
\end{eqnarray}
in the step 1 of our algorithm by the two-register state
\begin{eqnarray}
\label{Eq:20}
|\widehat{C}^{(k)}\rangle:=\frac{\sum_{j=1}^{M_c^{(k)}}|j\rangle|C_j^{(k)}\rangle}{\sqrt{M_c^{(k)}}}.
\end{eqnarray}
When $k=1$, $\mathcal{C}^{(k)}=\mathcal{I}$, i.e., $C^{(k)}_j=I_{j-1}$ and $M_c^{(k)}=M$, and we can efficiently create the state $|C^{(k)}\rangle=\frac{\sum_{j=0}^{M-1}|j\rangle}{\sqrt{M}}$ in time $\mathcal{O}(\log(M))$. But for the case when $k>1$, it is more desirable to use the state $|\widehat{C}^{(k)}\rangle$ instead of $|C^{(k)}\rangle$, because in most cases it is more efficient to create the former state than the latter one. To generate $|\widehat{C}^{(k)}\rangle$, a quantum oracle (denoted by $O_C$) that performs
\begin{eqnarray}
\label{Eq:21}
O_C\frac{\sum_{j=1}^{M_c^{(k)}}|j\rangle|0\rangle}{\sqrt{M_c^{(k)}}}=\frac{\sum_{j=1}^{M_c^{(k)}}|j\rangle|C_j^{(k)}\rangle}{\sqrt{M_c^{(k)}}},
\end{eqnarray}
is applied; this can be achieved via the quantum random access memory \cite{QRAM} in time $\mathcal{O}(k\log(MM_c^{(k)}))$ provided the classical data of candidate $k$-itemsets $C_j^{(k)}$. However, generating the state $|C^{(k)}\rangle$ from the initial $k$ $M$-dimensional states $|0\rangle^{\otimes k}$ by using amplitude amplification takes time $\mathcal{O}(k\log(M)\sqrt{\frac{M^k}{M_c^{(k)}}})$(creating $k$ uniform superposition states takes time $\mathcal{O}(k\log(M))$ and amplitude amplification takes $\mathcal{O}(\sqrt{\frac{M^k}{M_c^{(k)}}})$ repetitions), which in practice is much more time consuming than generating $|\widehat{C}^{(k)}\rangle$. It should be noted that, if the state $|\widehat{C}^{(k)}\rangle$ is taken in our algorithm, it is the state of the second register of the state, i.e., the mixed state $\frac{\sum_{j=1}^{M_c^{(k)}}|C_j^{(k)}\rangle\langle C_j^{(k)}|}{M_c^{(k)}}$ instead of the pure superposition state $|C^{(k)}\rangle$ (Eq. (19)), that will be operated in the step 2 and measured in the final step.

\subsection{Complexity analysis}
In the steps 1-3 of our algorithm, it takes  $T-1$ oracles $O^{(k)}$ and the error for estimating $s_j^{(k)}$ is $\Theta(\frac{\sqrt{s_j^{(k)}(1-s_j^{(k)})}}{T})$ \cite{AA}. Therefore, to ensure the error for estimating $s_j^{(k)}$ is $\epsilon\sqrt{s_j^{(k)}(1-s_j^{(k)})}$, $T$ should be taken as $T=\Theta(\frac{1}{\epsilon})$. In the step 4 for amplitude amplification, $\mathcal{O}(\sqrt{\frac{M_c^{(k)}}{M_f^{(k)}}})$ repetitions (iterations) are required. The last step takes $\mathcal{O}(M_f^{(k)})$ measurements to reveal all the $M_f^{(k)}$ frequent $k$-itemsets and their supports. Putting all of these quantities and noting that the construction of $O^{(k)}$ entails $\Theta(k)$ basic oracles $O$, our algorithm takes $\mathcal{O}(k\cdot T \cdot \sqrt{\frac{M_c^{(k)}}{M_f^{(k)}}} \cdot M_f^{(k)})=\mathcal{O}(\frac{k\sqrt{M_c^{(k)}M_f^{(k)}}}{\epsilon})$ basic oracles $O$ to mine all the $M_f^{(k)}$ frequent $k$-itemsets ($\mathcal{F}^{(k)}$) from $M_c^{(k)}$ candidate $k$-itemsets ($\mathcal{C}^{(k)}$) and estimate their supports.

Now we consider the classical sampling-based algorithm for mining $\mathcal{F}^{(k)}$ from $\mathcal{C}^{(k)}$, where the supports $s_j^{(k)}$ of all the candidate $k$-itemsets in $\mathcal{C}^{(k)}$ are estimated by sampling the transactions of the database $\mathcal{T}$. According to the properties of Binomial distribution, to ensure the induced error $\epsilon\sqrt{s_j^{(k)}(1-s_j^{(k)})}$ for estimating $s_j^{(k)}$, it needs $\mathcal{O}(\frac{1}{\epsilon^2})$ samples to estimate every support (same number of samples is used to estimate every support). The errors are with the same scales as those in our quantum algorithm. Since $\Theta(k)$ basic oracles $O$ are required to identify whether a certain sample (transaction) contains an arbitrary $k$-itemset $X$, it will take $\mathcal{O}(\frac{kM_c^{(k)}}{\epsilon^2})$ basic oracles $O$ to estimate all the supports of $M_c^{(k)}$ candidate $k$-itemsets in $\mathcal{C}^{(k)}$ with precision $\mathcal{O}(\epsilon)$. After estimating the supports by sampling, one can easily find out the frequent $k$-itemsets and obtain their supports.

However, both of above two algorithms are non-deterministic. If we want to mine $\mathcal{F}^{(k)}$ from $\mathcal{C}^{(k)}$ in a deterministic way, we can directly take the classical Apriori algorithm. In the algorithm, every transaction of database is scanned to calculate the support of every candidate $k$-itemsets, and thus $\mathcal{O}(kM_c^{(k)}N)$ basic oracles $O$ are required to calculate all the supports of $M_c^{(k)}$ candidate $k$-itemsets and no errors are induced at all. After calculation, one can directly find out the frequent $k$-itemsets and obtain their supports.

The comparison of our algorithm, classical sampling-based algorithm and the Apriori algorithm for mining $\mathcal{F}^{(k)}$ from $\mathcal{C}^{(k)}$ is given in the Table \ref{tab:1}. From the comparison, two points are derived. First, our quantum algorithm and the classical sampling-based algorithm are more efficient than the Apriori algorithm when the number of transactions $N$ is large in most cases, while these two algorithms are non-deterministic and induce errors. So there is a trade-off between the accuracy and the complexity. Second, more importantly, compared with the query complexity of classical sampling-based algorithm, the query complexity of our quantum algorithm quadratically improves the dependence on the error. Since  $M_f^{(k)} \leq M_c^{(k)}$, the improvement in the dependence on the parameter $M_c^{(k)}$ is also achieved, but the degree relies on the scale of $M_f^{(k)}$ relative to $M_c^{(k)}$ . When $M_f^{(k)} \approx M_c^{(k)}$, no significant improvement is achieved. However, when $M_f^{(k)}\ll M_c^{(k)}$, quadratic improvement in the dependence on $M_c^{(k)}$ is also achieved. It is conceivable that this situation probably happens in the last iteration with setting a high minimum support threshold, because (1) the number of frequent itemsets in the last iteration would be too small to generate candidate itemsets for the next iteration; (2) higher threshold implies smaller number of frequent itemsets existing in candidate itemsets.
\begin{table}[b]
\caption{\label{tab:1}%
Comparisons of our quantum algorithm, classical sampling-based algorithm and the classical Apriori algorithm for mining $\mathcal{F}^{(k)}$ from $\mathcal{C}^{(k)}$.}
\begin{ruledtabular}
\begin{tabular}{ccc}
algorithm & determinacy & query complexity \\
\hline
Quantum          &non-deterministic  &$\mathcal{O}(\frac{k\sqrt{M_c^{(k)}M_f^{(k)}}}{\epsilon})$ \\
Sampling-based   &non-deterministic  &$\mathcal{O}(\frac{kM_c^{(k)}}{\epsilon^2})$ \\
Apriori          &deterministic      &$\mathcal{O}(kM_c^{(k)}N)$
\end{tabular}
\end{ruledtabular}
\end{table}

Regarding the complexity of our algorithm, another two issues should also be addressed.

\emph{1. The overall query complexity of our algorithm for mining all the frequent itemsets.}

Since our quantum algorithm together with all the classical ARM algorithms finally output all the frequent itemsets instead of frequent $k$-itemsets for one particular $k$, it is necessary to analyze the overall query complexity for generating all the frequent itemsets. Assuming $\widehat{k}$ iteration are performed in the algorithm, the overall query complexity would be $\mathcal{O}(\frac{\sum_{k=1}^{\widehat{k}}k\sqrt{M_c^{(k)}M_f^{(k)}}}{\epsilon})$, while the overall complexity of the classical sampling-based algorithm is $\mathcal{O}(\frac{\sum_{k=1}^{\widehat{k}}kM_c^{(k)}}{\epsilon^2})$. Just as mining frequent $k$-itemsets shown above, the improvement of our algorithm over the classical algorithm is also consist of two parts. First, the quadratic improvement on $\epsilon$ contributed by parallel amplitude estimation is conclusive. Second, however, the improvement contributed by amplitude amplification depends on the database itself and the threshold $min\_supp$ because these two determine the sizes of $M_c^{(k)}$ and $M_f^{(k)}$. To quantify the improvement caused by amplitude amplification, we take the value
\begin{eqnarray}
\label{Eq:22}
\gamma:=\frac{\sum_{k=1}^{\widehat{k}}kM_c^{(k)}}{\sum_{k=1}^{\widehat{k}}k\sqrt{M_c^{(k)}M_f^{(k)}}},
\end{eqnarray}
 which means our algorithm is roughly $\gamma$ times faster than the classical algorithm (regardless of the improvement caused by parallel amplitude estimation), as a measure. To show $\gamma$ depends on the database itself and the threshold, two real-world transaction databases, retail and kosarak \cite{FIM}, which are usually taken to test the classical ARM algorithms, are ran by Apriori algorithm with each one taking two thresholds, 1\% and 2\%; see the appendix for details. By simple calculation, we derive $\gamma=$12.75, 25.54, 19.87 and 33.74 for the four cases, (retail, 1\%), (retail, 2\%), (kosarak, 1\%) and (kosarak, 2\%), respectively.

\emph{2. The overall time complexity of invoking the operation of generating the states $|C^{(k)}\rangle$ and $|\widehat{C}^{(k)}\rangle$.}

According to the last paragraph of last subsection, we know that in our algorithm the states $|C^{(k)}\rangle$ and $|\widehat{C}^{(k)}\rangle$ need to be prepared for $k=1$ and $k>1$ repectively, and each one of the two states can be generated in time $\mathcal{O}(k\log(MM_c^{(k)}))$ (incorporating the case $k=1$). Taking amplitude amplification in step 4 and measurement in step 5 into consideration, the overall time complexity of invoking the operation of generating $|C^{(k)}\rangle$ and $|\widehat{C}^{(k)}\rangle$ would be $\mathcal{O}(\log(MM_c^{(k)})k\sqrt{M_c^{(k)}M_f^{(k)}})$. Compared with the overall query complexity of calling the basic oracles $O$, $\mathcal{O}(\frac{k\sqrt{M_c^{(k)}M_f^{(k)}}}{\epsilon})$, the overall time complexity of the invocation of the operation of generating the states $|C^{(k)}\rangle$ and $|\widehat{C}^{(k)}\rangle$ is less costly. This is because in practice $\log(MM_c^{(k)})$ is much smaller than $\frac{1}{\epsilon}$ especially when $\epsilon$ is set to be very small ($\epsilon=0.001$ for example). That is to say, it is the time complexity of calling the basic oracle $O$ that dominates the overall time complexity of our quantum algorithm.
\section{Conclusion}
In this paper, we address ARM, one of the most important problems in data mining, in the quantum settings. We provide a quantum algorithm for the core procedure of implementing ARM, mining frequent itemsets from the candidate itemsets. Specifically, by subtly using amplitude estimation and amplitude amplification, our algorithm can efficiently find out the frequent $k$-itemsets from candidate $k$-itemsets and estimate their supports. Complexity analysis shows our algorithm is faster than the classical counterpart, classical sampling-based algorithm, in the sense that the complexity of our algorithm is at least quadratically improved in the dependence on the error. We hope our quantum algorithm for ARM can help in better understanding the power of quantum computing and inspire more quantum algorithms for big data mining tasks.

In the future, we will further investigate quantum ARM from two aspects. First, noting that our algorithm in this paper focuses on efficiently implementing the procedure (P1) mentioned in Sec. \ref{sec:2}, quantum algorithms for the procedure (P2) should be explored. Second, it is interesting to introduce privacy protection into quantum ARM. A recent work on this topic has been put forward in \cite{YYF}.

\section*{Acknowledgements}
We would like to thank the anonymous referees for their
helpful comments. This work is supported by NSFC (Grant Nos. 61272057, 61572081).

\appendix*
\section{The details of running classical Apriori algorithm on two real-world transaction databases}
We run the Apriori algorithm \cite{Apriori} on two real-world transaction databases, retail and kosarak, which contain 88162 transactions and 16470 items, and 992547 transactions and 41270 items repectively. We obtain the the numbers of candidate itemsets and frequent itemsets in each iteration, which is shown in the Table \ref{tab:table2} and Table \ref{tab:table3}, for two minimum support thresholds, 1\% and 2\%. While five iterations are executed for kosarak for both thresholds, four iterations are executed for retail for both thresholds.
\begin{table}[h]
\caption{\label{tab:table2}%
The numbers of candidate itemsets and frequent itemsets, $M_c^{(k)}$ and $M_f^{(k)}$, for the database retail. Here $k$ labels the $k$th iteration.}
\begin{ruledtabular}
\begin{tabular}{ccccc}
\multirow{2}{*}{$k$} &\multicolumn{2}{c}{$min\_supp=1\%$} &\multicolumn{2}{c}{$min\_supp=2\%$}\\
\cline{2-3} \cline{4-5}
 &$M_c^{(k)}$ &$M_f^{(k)}$ &$M_c^{(k)}$ &$M_f^{(k)}$ \\ \hline
1  &16470 &70 &16470 &20 \\
2 &2415 &58 &190 &22 \\
3 &37 &25 &14 &12\\
4 &6 &6 &2 &1
\end{tabular}
\end{ruledtabular}
\end{table}

\begin{table}[h]
\caption{\label{tab:table3}%
The numbers of candidate itemsets and frequent itemsets for the database kosarak.}
\begin{ruledtabular}
\begin{tabular}{ccccc}
\multirow{2}{*}{$k$} &\multicolumn{2}{c}{$min\_supp=1\%$} &\multicolumn{2}{c}{$min\_supp=2\%$}\\
\cline{2-3} \cline{4-5}
 &$M_c^{(k)}$ &$M_f^{(k)}$ &$M_c^{(k)}$ &$M_f^{(k)}$ \\ \hline
1  &41270 &54 &41270 &27 \\
2 &1431 &140 &351 &45 \\
3 &194 &127 &45 &34\\
4 &57 &52 &13 &13\\
5 &11 &10 &2 &2
\end{tabular}
\end{ruledtabular}
\end{table}


\begin{thebibliography}{ss3}
\bibitem{Complexity} $\mathcal{O}(f(n))$ and $\Omega(f(n))$ repectively refer to functions that are $\leq c_1f(n)$ and $\geq c_2f(n)$ for some cases of $c_1,c_2>0$ and sufficiently large $n$. $\Theta(f(n))$ refers to a fuction that is both $\mathcal{O}(f(n))$ and $\Omega(f(n))$.
\end{thebibliography}
\end{document}